\documentclass[acmsmall]{acmart}

\copyrightyear{2025}
\acmYear{2025}
\acmDOI{XXXXXXX.XXXXXXX}

\acmConference[Conference FSE'26]{Montreal, Canada}
\acmISBN{978-1-4503-XXXX-X/2025/08}

\usepackage[dvipsnames, svgnames, x11names, table]{xcolor}
\usepackage{colortbl}%
\usepackage{easyReview}
 \usepackage[UTF8, scheme=plain, punct=plain, zihao=false]{ctex}
\usepackage{hyperref}
\usepackage{subcaption}
\usepackage{graphicx}
\usepackage{epsfig}

\usepackage{algorithm}
\usepackage{algpseudocode}
\usepackage{algorithmicx}
\algrenewcommand\algorithmicrequire{\textbf{Input:}}
\algrenewcommand\algorithmicensure{\textbf{Output:}}

\usepackage[group-separator={,},group-minimum-digits=4]{siunitx}

\usepackage{tikz}
\newcommand{\pgftextcircled}[1]{
    \setbox0=\hbox{#1}%
    \dimen0\wd0%
    \divide\dimen0 by 2%
    \begin{tikzpicture}[baseline=(a.base)]%
        \useasboundingbox (-\the\dimen0,0pt) rectangle (\the\dimen0,1pt);
        \node[circle,draw,outer sep=0pt,inner sep=0.1ex] (a) {#1};
    \end{tikzpicture}
}
\let\textcircled=\pgftextcircled

\usepackage{listings}
\usepackage{pifont}
\usepackage{enumitem}
\usepackage{multirow} 
\usepackage{caption}
\usepackage[framemethod=TikZ]{mdframed}
\usepackage{siunitx}
\usepackage{tcolorbox}
\usepackage{float}  
\usepackage{makecell}
\usepackage{algpseudocode}

\definecolor{mygray}{gray}{.9}
\definecolor{mycyan}{cmyk}{.3,0,0,0}
\definecolor{mauve}{rgb}{0.58,0,0.82}
\definecolor{dkgreen}{rgb}{0,0.6, 0}
\definecolor{gray}{rgb}{0.5,0.5,0.5}
\DeclareCaptionFormat{mylst}{\hrule#1#2#3}
\newcommand{\methodName}{PyTIR}

\author{Shuo Sun}
\affiliation{%
   \institution{Technology Center of Software Engineering, Institute of Software, Chinese Academy of Sciences}
   \city{Beijing}
   \country{China}}
\email{sunshuo20@otcaix.iscas.ac.cn}

 \author{Shixin Zhang}
\affiliation{%
   \institution{Technology Center of Software Engineering, Institute of Software, Chinese Academy of Sciences}
   \city{Beijing}
   \country{China}}
\email{zhangshixin@otcaix.iscas.ac.cn}

 \author{Jiwei Yan}
\affiliation{%
   \institution{Technology Center of Software Engineering, Institute of Software, Chinese Academy of Sciences}
   \city{Beijing}
   \country{China}}
\email{ yanjiwei@otcaix.iscas.ac.cn}

 \author{Jun Yan}
\affiliation{%
  \institution{Technology Center of Software Engineering, Institute of Software, Chinese Academy of Sciences}
   \city{Beijing}
   \country{China}}
\email{ yanjun@ios.ac.cn}

 \author{Jian Zhang}
\affiliation{%
  \institution{Key Laboratory of System Software (Chinese Academy of Sciences), Institute of Software, Chinese Academy of Sciences}
   \city{Beijing}
   \country{China}}
\email{zj@ios.ac.cn}

\begin{document}

\title{Co-Evolution of Types and Dependencies: Towards Repository-Level Type Inference for Python Code}

\begin{abstract}

Python's dynamic typing mechanism, while promoting flexibility, is a significant source of runtime type errors that plague large-scale software, which inspires the automatic type inference techniques.
Existing type inference tools have achieved advances in type inference within isolated code snippets.
However, repository-level type inference remains a significant challenge, primarily due to the complex inter-procedural dependencies that are difficult to model and resolve.
To fill this gap, we present \methodName, a novel approach based on LLMs that achieves repository-level type inference through the co-evolution of types and dependencies. 
\methodName~constructs an Entity Dependency Graph (EDG) to model the objects and type dependencies across the repository. During the inference process, it iteratively refines types and dependencies in EDG for accurate type inference. Our key innovations are: (1) an EDG model designed to capture repository-level type dependencies; (2) an iterative type inference approach where types and dependencies co-evolve in each iteration; and (3) a type-checker-in-the-loop strategy that validates and corrects inferences on-the-fly, thereby reducing error propagation.
When evaluated on 12 complex Python repositories, \methodName~significantly outperformed prior works, achieving a \textit{TypeSim} score of 0.89 and a \textit{TypeExact} score of 0.84, representing a 27\% and 40\% relative improvement over the strongest baseline. More importantly, \methodName~removed new type errors introduced by the tool by 92.7\%. This demonstrates a significant leap towards automated, reliable type annotation for real-world Python development. 

\end{abstract}
\maketitle
\section{Introduction}

Python's \textit{dynamic typing} provides substantial flexibility and development speed, making it especially advantageous for rapidly building data processing scripts, a key factor in its widespread adoption~\cite{githubOct, IEEESpectrum}. However, as Python has evolved from a scripting language into a primary development choice for large-scale repositories~\cite{alfadel2023empirical,yang2022complex, rak2020python}, the flexibility introduces potential risks. In particular, the absence of compile-time type checking can create latent issues that may remain undetected until runtime, potentially impacting system reliability.
To mitigate this issue, PEP~484 introduced a standardized syntax for \textit{type annotations}~\cite{PEP484} in Python, enabling static type checkers, linters, and IDEs to detect potential errors before execution. 
Nevertheless, writing type annotations manually can be labor-intensive for developers. 
A previous study~\cite{rak2020python} revealed a notably low usage rate, finding that only 2,678 out of 70,000 analyzed repositories had adopted type annotations.
This observation underscores the pressing need for automatic type inference techniques to alleviate the burden of manual annotation.


In recent years, many studies have explored automatic type inference techniques for Python code, encompassing both \textit{static-analysis-based (SA-based)} \cite{wu_quac_2024,peng_static_2022} and \textit{learning-based} \cite{peng_generative_2023,wang_tiger_2024} approaches.
On the one hand, \textit{SA-based} approaches leverage pre-defined inference rules, which enable efficient inference with high accuracy in restricted scenarios.
On the other hand, \textit{learning-based} approaches, ranging from deep learning models to Large Language Models (LLMs), offer greater flexibility by exploiting code comprehension capabilities and Python-specific syntactic knowledge, which can achieve high accuracy when sufficient code context is available.
For both types, the empirical evaluations on standardized benchmarks, primarily consisting of isolated code snippets, demonstrate the effectiveness.
For example, on the widely used \textit{ManyTypes4Py} benchmark~\cite{mir2021manytypes4py}, existing methods achieve an accuracy of up to 86.2\%~\cite{wang_tiger_2024}, indicating the potential of current techniques in practical type inference tasks.
However, when applied to real-world Python repositories, which often involve complex dependencies among Python objects, existing approaches face significant challenges.
As illustrated in our preliminary study (see motivating examples in Section~\ref{Motivating}), both \textit{SA-based} and \textit{learning-based} approaches, as well as native LLMs, frequently fail to infer correct types in the presence of complex dependencies, particularly inter-procedural dependencies, which are commonly encountered in real-world codebases.
This limitation underscores a persistent gap between current research and its applicability to real-world Python repositories.


To fill this gap, we aim to propose a \textit{practical} and \textit{precise} \textbf{repository-level Python type inference} approach.
Leveraging the superior semantic understanding capabilities of LLMs, we adopt LLMs as the core inference engine of our approach.
Based on our investigation, we have identified three key challenges to our approach.

\textbf{C1: Identifying and resolving complex \textit{inter-procedural type dependencies}.}
In real-world Python repositories, objects are interconnected through both intra-procedural and inter-procedural type dependencies. 
In contrast to intra-procedural type dependencies, where all relevant variables exist within a local context, resolving inter-procedural type dependencies presents greater challenges.
The objects involved in inter-procedural dependencies are often dispersed across various modules within the repository, making them difficult to locate.
Moreover, Python's dynamic binding mechanism further complicates their identification. 
However, these inter-procedural dependencies frequently provide critical type inference information. 
Therefore, a key challenge lies in accurately identifying and extracting these dependencies, as well as effectively incorporating them into the type inference process.

\textbf{C2: Extracting compact yet highly relevant \textit{type-inference contexts} for LLMs.}
On code understanding and analysis tasks, LLMs have already demonstrated remarkable capabilities.
However, even with increasingly large context windows, it is still impractical to provide an entire code repository as input. 
Moreover, supplying excessive information at once may distract the model's attention and negatively affect inference quality.
Thus, it is essential to construct \textit{type-inference contexts} that are both compact and highly relevant to the target object. 
The difficulty lies in that naive context selection strategies, such as those based solely on distance, risk discarding critical information, which in turn reduces inference precision.
Thus, a key challenge lies in abstracting and constructing proper contextual information that enables LLM-based analysis to obtain sufficient knowledge.


\textbf{C3: Preventing \textit{incorrect types} from propagating along dependency chains.}
Due to the stochastic nature of LLMs and potential hallucinations, type inference results may contain inaccuracies.
Without timely correction, an incorrectly inferred type at any point in a dependency chain can propagate to subsequent dependencies, adversely affecting downstream type inference.
Thus, a key challenge lies in designing real-time error detection and corrective feedback mechanisms capable of identifying and rectifying type inconsistencies before these inaccuracies.

To address these challenges, we propose \textbf{\methodName} (\textbf{Py}thon \textbf{T}ype \textbf{I}nference on \textbf{R}epository-Level), a novel approach for practical and precise repository-level Python type inference.
To address \textbf{Challenge 1}, \methodName~introduces a data structure called the \textbf{Entity Dependency Graph (EDG)}, which captures essential type dependencies. 
We further propose an iterative approach for repository-level EDG construction. 
The core idea is \textit{Co-Evolution of Types and Dependencies}. In each iteration, the dependencies are updated based on the latest inferred type information, which in turn refines the EDG model and guides subsequent type inference.
To address \textbf{Challenge~2}, \methodName~transforms the EDG into a directed acyclic graph and performs analysis following a topological ordering. 
After reorganization, \methodName~infers nodes with fewer dependencies first, and summarizes their type information as contextual input for dependent nodes. Moreover, nodes that mutually depend on each other are treated as a cluster, i.e., their contexts are combined, and \methodName~infers their types simultaneously.
Finally, to address \textbf{Challenge 3}, \methodName~leverages a static type checker to detect potential type conflicts within the repository after each inference iteration. 
When a conflict is identified, a heuristic strategy is applied to locate the erroneous type annotations and fix them.



We evaluate the effectiveness of \methodName~on 12 large-scale real-world Python repositories.
The evaluation measures both the accuracy of the generated type annotations and the number of type errors introduced by the type annotations.
In terms of type accuracy, \methodName~outperforms all baseline methods, achieving a top score of 0.89 on \textit{TypeSim} metric and 0.84 on \textit{TypeExact} metric, corresponding to improvements of 27\% and 40\% over the previous state-of-the-art (SOTA) methods.
With respect to introduced type errors, \methodName~reduces 92.7\% of introduced type errors compared to the SOTA methods, resulting in an average of only 46.17 errors per repository.


To conclude, the contributions of this paper are as follows:
\begin{itemize}
\item We design a novel data structure, the Entity Dependency Graph (EDG), to model type dependencies for repository-level Python type inference.
\item We propose an automated type inference approach for Python repositories based on the key idea, \textit{Co-Evolution of Types and Dependencies}, achieving high accuracy while reducing type errors introduced.
\item We implement our approach into an open-source tool \methodName; we also provide a high-quality dataset comprising 12 unannotated Python repositories and their corresponding type-annotated versions, which are free of type errors and can serve as benchmarks for further evaluation. 
\end{itemize}

\section{Background and Motivating Example}

In this section, we first describe the various type dependencies and then illustrate the challenges of repository-level Python type inference through motivating examples.

\subsection{Type Dependencies in Python}

The core insight of repository-level type inference is that the types of program entities are not independent but are constrained by a complex web of relationships. 
Accurately modeling these dependencies is crucial for propagating type information and achieving high inference accuracy.
As pointed out by prior work~\cite{peng_static_2022}, type dependencies in Python exist between \textit{variable occurrences} rather than \textit{variables} themselves. 
It is due to Python's dynamic typing nature, where a variable can be rebound to values of different types during runtime.
In this paper, we call a variable occurrence $o_1$ \textit{dependent} on another variable occurrence $o_2$, denoted as $o_1 \rightarrow o_2$, if the type of $o_2$ is a prerequisite for determining the type of $o_1$.
The common type dependencies can be categorized as follows:
\begin{itemize}
\item \textit{Usage-Definition Dependency}: In Python, a variable can be reassigned to a value of a different type, and the type of a variable at a usage occurrence is determined by its most recent preceding definition occurrence. Therefore, the usage occurrence of a variable is \textit{dependent} on its most recent preceding definition occurrence. 
\item \textit{Parameter-Usage Dependency}: Due to the principle of duck typing, the set of valid types for a function parameter is constrained by how that parameter is used within the function body. Therefore, a parameter occurrence is \textit{dependent} on all its usage occurrences within the function.
\item \textit{Argument-Parameter Dependency}: The type of an argument in a function call must be compatible with the type of the corresponding parameter. Therefore, a call-site argument occurrence is \textit{dependent} on the function's parameter occurrence.
\item \textit{Attribute-Instance Dependency}: The type of an attribute access, such as \texttt{instance.attr}, is determined by the type of the \texttt{instance} object. Therefore, the attribute occurrence is \textit{dependent} on the instance occurrence.
\item \textit{Semantic Dependency}: This category covers dependencies derived from the operational semantics of the code, rather than explicit syntactic links. 
For instance, the expression \texttt{x+y} creates a semantic dependency between \texttt{x} and \texttt{y}, as each variable constrains the possible types of the other through the semantics of the `+' operator.
\end{itemize}

Furthermore, these dependencies can be categorized based on their scope.
\textit{Intra-procedural dependencies} are those resolved within the scope of a single function. 
In contrast, \textit{inter-procedural dependencies} span across function boundaries. 
Resolving inter-procedural dependencies is more challenging as it requires tracking type flows across function calls, often necessitating a whole-program analysis approach.

\subsection{Motivating Example}~\label{Motivating}



\begin{figure}[t!]
    \centering
    \begin{subfigure}[b]{\textwidth}
        \centering
        \resizebox{0.95\columnwidth}{!}{
            \includegraphics[width=\textwidth]{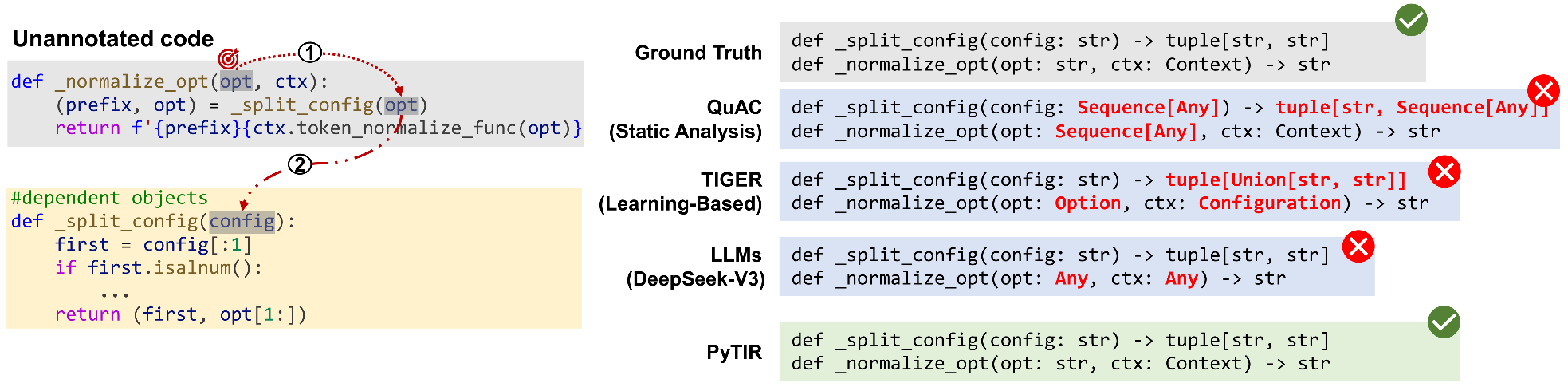}
        }
        \caption{Code Snippet with \textit{Simple} Inter-Procedural Dependencies}
        \label{fig:simple_dep}
    \end{subfigure}
    
    
 \begin{subfigure}[b]{\textwidth}
        \centering
        \resizebox{0.95\columnwidth}{!}{
            \includegraphics[width=\textwidth]{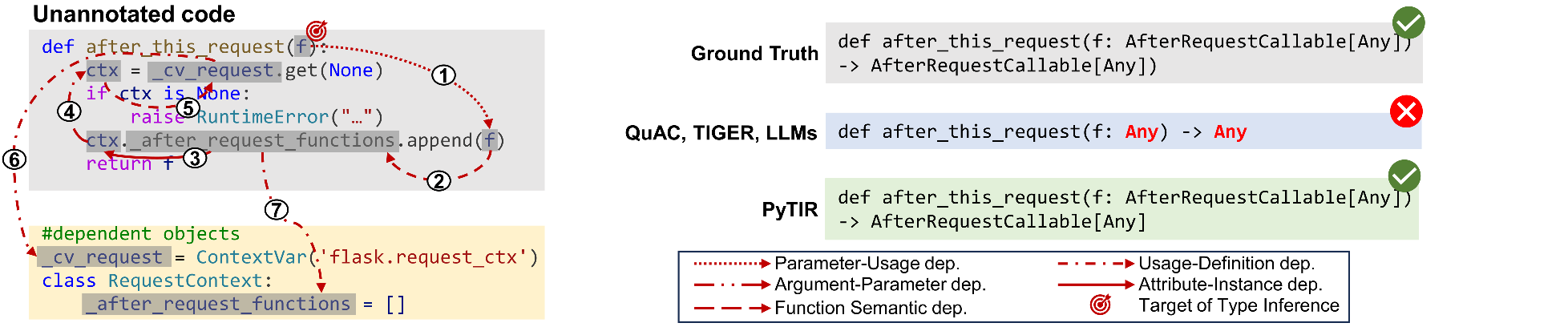} 
        }
        \caption{Code Snippet with \textit{Complex} Inter-Procedural Dependencies}
        \label{fig:complex_dep}
    \end{subfigure}
    \vspace{-1.5em}
    \caption{Motivating Examples and Their Type Inference Results}
    \vspace{-1.5em}
    \label{fig:motivating_example}
\end{figure}

In the following, we use two real-world examples to illustrate our motivation.
Fig.~\ref{fig:simple_dep} and Fig.~\ref{fig:complex_dep} show two code snippets collected from the widely used Python repositories \textit{Click}~\cite{click} and \textit{Flask}~\cite{flask}. 
When applying SOTA SA-based approaches (\textit{QuAC}~\cite{wu_quac_2024}), learning-based approaches (\textit{TIGER}~\cite{wang_tiger_2024}), and modern LLMs (\textit{DeepSeek-V3~\cite{liu2024deepseek}}, \textit{DeepSeek-R1}~\cite{guo2025deepseek},\textit{GPT-4o}~\cite{hurst2024gpt}) for type inference, all methods fail to produce correct type annotations.
The primary reason for this phenomenon lies in the various types of dependencies present in real-world repositories.
By further analysis, we reveal that these approaches consistently overlook three critical aspects: (1) accurately identifying and extracting inter-procedural dependencies, (2) timely correcting erroneous type inferences due to dependency missing, and (3) iteratively co-evolving dependencies and types.
Furthermore, we will provide a detailed comparison of the type inference results produced by different methods.

\textbf{(1) Dependency Identification and Extraction.}
Fig.~\ref{fig:simple_dep} illustrates the type inference process of the parameter \texttt{opt} in function \texttt{\_normalize\_opt} by tracing type dependencies.
To determine the type of the parameter \texttt{opt}, the analysis must first examine its usages within the function body. 
As indicated by dependency \textcircled{1}, a key usage of \texttt{opt} is as an argument in the call to the function \texttt{\_split\_config}. 
It establishes an intra-procedural dependency between the parameter declaration and its application at the call site.
Subsequently, the type constraints imposed on the argument \texttt{opt} are propagated from the corresponding parameter \texttt{config} in the callee function \texttt{\_split\_config}. This inter-procedural dependency, denoted by \textcircled{2}, demonstrates that inferring the type of the argument \texttt{opt} is contingent upon the type of the parameter \texttt{config}. 
The complete type inference for \texttt{opt} in \texttt{\_normalize\_opt} requires resolving this dependency chain, which necessitates analyzing the callee function \texttt{\_split\_config}.
However, neither \textit{TIGER} nor modern \textit{LLMs} captured the inter-procedural dependency \textcircled{2}, consequently failing to infer the correct type for parameter \texttt{opt} in function \texttt{\_normal\_opt}.

To address this challenge, \methodName~designs a data structure EDG to model type dependencies and employs an iterative algorithm to extract them.
As shown in Fig.~\ref{fig:simple_dep}, \methodName~correctly identifies the inter-procedural dependency and infers the correct type  for \texttt{\_normalize\_opt}.

\textbf{(2) Error Detection and Correction.} 
Although QuAC correctly identified the inter-procedural dependency \textcircled{2} in Fig.\ref{fig:simple_dep}, it still failed to infer the correct type of \texttt{opt}.
This incorrect inference stems from the wrong type of \texttt{config} in the function \texttt{\_split\_config}, on which \texttt{opt} depends.
This error type then propagates through the dependency chain \textcircled{1}-\textcircled{2}, causing further error types.

To overcome this challenge, \methodName~incorporates a static type checker to identify and correct erroneous type annotations immediately after generation, preventing further error propagation through dependency chains. 
Therefore, once \methodName~also generates a wrong parameter type in the function \texttt{\_split\_config}, \methodName~can detect the resulting type conflicts and fix the wrong type according to the error message from the static type checker.

\textbf{(3) Co-evolution of Types and Dependencies.} 
Moreover, while the demonstrated call dependency in Fig.\ref{fig:simple_dep} is relatively straightforward, real-world repositories also involve far more intricate dependencies, which significantly hinder the accuracy of type inference.
For example, Fig.~\ref{fig:complex_dep} shows a more complex dependency chain \textcircled{1}-\textcircled{7} to infer the type of parameter \texttt{f} in function \texttt{after\_this\_request}.
This chain reveals that inferring the type of \texttt{f} is contingent upon determining the type of \texttt{ctx.\_after\_request\_functions} (dependencies \textcircled{1}-\textcircled{2}).
Therefore, the usage-definition dependency\textcircled{7} between the usage occurrence and definition occurrence of  \texttt{ctx.\_after\_request\_functions} is crucial for inferring the type of \texttt{f}.
However, due to Python's dynamic nature, without the specific type of \texttt{ctx}, the object that \texttt{ctx.\_after\_request\_functions} refers to cannot be statically resolved.
Consequently, existing type inference methods QuAC, TIGER, and native LLMs all failed to identify dependency \textcircled{7} and are thus unable to infer the type of \texttt{f}.

To resolve this challenge, \methodName~employs a co-evolution process where the dependencies and
types mutually refine each other. 
Specifically, \methodName~first infered the type for the local variable \texttt{ctx} using dependences \textcircled{4}\textcircled{5}\textcircled{6} in the incomplete dependency chain \textcircled{1}-\textcircled{6}.
Leveraging this newly acquired type information, it then identified the previously missing dependency \textcircled{7}, thereby completing the entire dependency chain. 
Finally, with the complete chain (\textcircled{1}-\textcircled{7}) established, \methodName~ successfully infers the precise type of \texttt{f}.

\section{Repository-level Python Type Inference Approach}
This section introduces our solution for ``\textit{how to perform practical and precise repository-level Python type inference}''.

\subsection{Approach Overview}

\begin{figure}[h!]
    \vspace{-10pt}
    \centering
    \resizebox{0.95\columnwidth}{!}{
        \includegraphics[width=\linewidth]{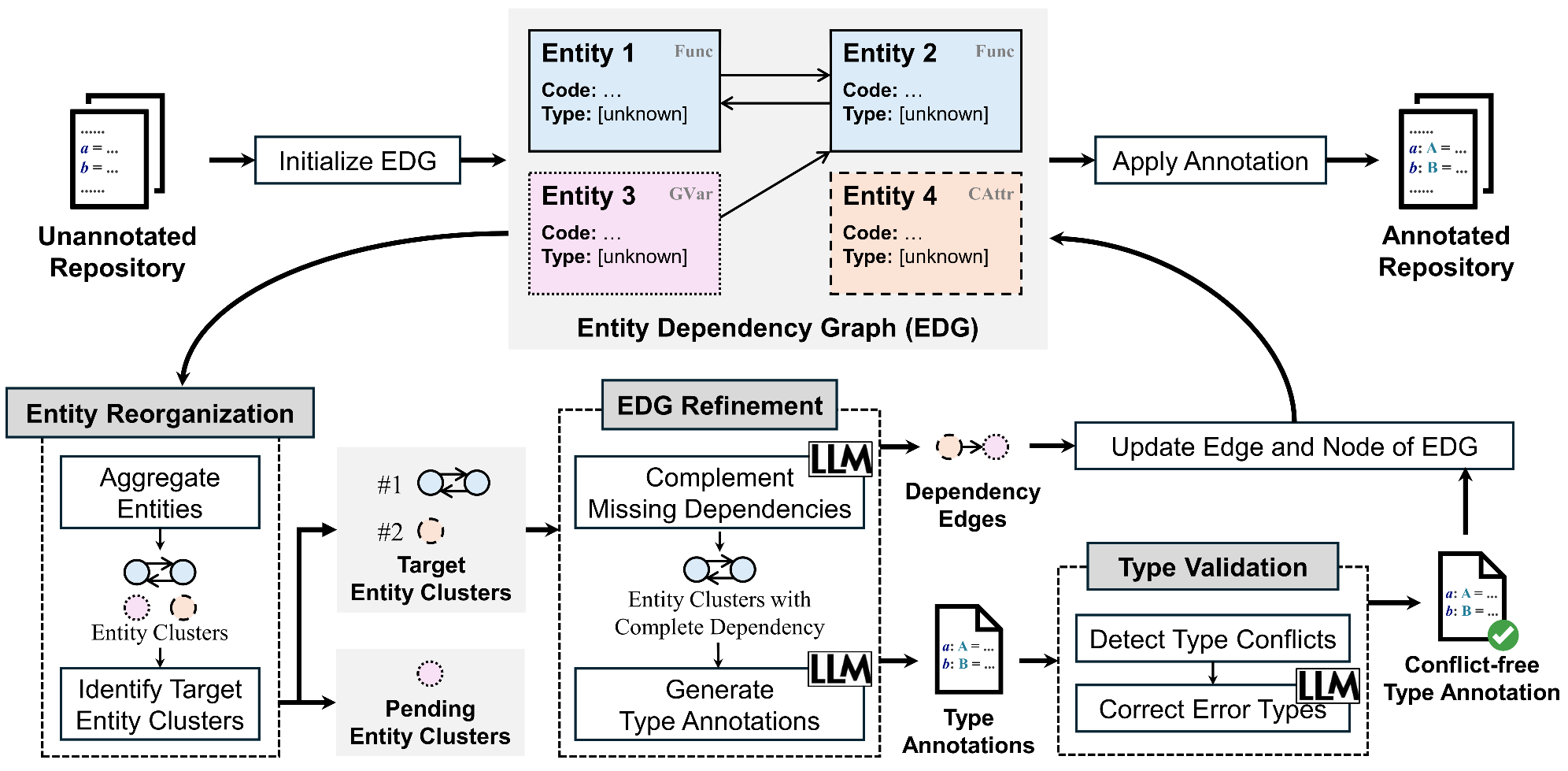}
    }
    \vspace{-10pt}
    \caption{Overview of \methodName~'s Workflow}
    \vspace{-8pt}
    \label{fig:method_overview}
\end{figure}

Fig.~\ref{fig:method_overview} illustrates the workflow of \methodName~ for repository-level type inference. Given a Python repository with incomplete type annotations (i.e., type hints), \methodName~ constructs an Entity Dependency Graph (EDG) to model both the objects requiring annotation and type dependencies in the repository. 
This graph is subsequently refined through an iterative process and ultimately applied to the original repository to produce a fully type-annotated version. The refinement iteration consists of three major phases.

\textbf{(1) Entity Reorganization}. To overcome the input size limitation of LLMs, \methodName~ reorganizes the complete EDG into entity clusters, which serve as separate type-inference contexts provided to the LLM. 
Correspondingly, the dependency edges in the EDG are also remanaged into dependencies between entity clusters. 
Then, \methodName~selects the target entity clusters based on the topological order of these dependencies, resulting in a sequence of entity clusters to be analyzed.

\textbf{(2) EDG Refinement}. 
In this phase, \methodName~ analyzes the target entity clusters to refine the dependencies and types in the EDG. 
\methodName~ first checks whether the dependencies in the EDG are sufficient for type inference for each entity cluster, and
identifies the lacking dependencies.
If an entity cluster's dependencies are complete and all dependent entities are annotated, type inference will be performed on this entity cluster, with the help of LLMs.

\textbf{(3) Type Validation}. To timely detect type errors and prevent their propagation along dependency chains, \methodName~ checks the generated type annotations with a static type checker.
Once a type conflict is detected, \methodName~leverages a backtracking algorithm to search for an appropriate type to resolve the conflict. 
If \methodName~fails to find a suitable type, it replaces the generated type annotation with `\texttt{Any}', sacrificing some type precision to ensure that no type conflicts exist in the generated type annotations.

These three phases will be performed iteratively.
In each iteration, the missing edges identified in the \textit{EDG Refinement} Phase and the conflict-free type annotations generated in the \textit{Type Validation} Phase are both updated into the EDG. 
When the type annotations for all nodes in the EDG have been inferred, the iteration terminates. 
\methodName~ then applies the type annotations from the EDG to the repository, ultimately producing a fully annotated repository.

\subsection{Design of Entity Dependency Graph}
As stated in previous sections, the intra- and inter-procedural type dependencies are crucial for repository-level type inference. 
Thanks to their remarkable capabilities in code comprehension, LLMs can effectively identify intra-procedural type dependencies when provided with the source code of a single function. 
To extend this reasoning capability to encompass inter-procedural dependencies that span multiple functions, classes, and files, \methodName~ transforms the entire repository into a structured, global representation, the Entity Dependency Graph (EDG). 
The EDG is a directed graph represented by a 2-tuple $G = \langle V, E \rangle$, where $V$ is the set of \textit{entities}, representing the objects that possess repository-level visibility, and $E \subseteq V \times V$ is the set of \textit{inter-entity dependencies}, which model the inter-procedural type dependencies in the repository.
In the following, we provide a comprehensive elaboration on \textit{entities} and \textit{inter-entity dependencies}, and detail the construction methodology of the EDG.

\subsubsection{Entities}

To model the inter-procedural type dependencies in the repository, \methodName~creates nodes in the EDG for objects whose types have cross-functional or global significance, which we call \textbf{Entities}.
The entities we consider are as follows.

\begin{itemize}
    \item \textbf{Variable Entities} include global variables defined at the module level and class attributes defined within classes.
    \item \textbf{Function Entities} include functions defined at the module, and class methods defined with classes. Specifically, the types of their parameters and return values will be annotated.
    \item \textbf{Class Entities} include all user-defined classes. These classes need no type annotation but can provide essential context for the type inference of other entities.
\end{itemize}

\methodName~does not explicitly target the inference of types for local variables within functions or for nested functions, which are typically confined to a local context. 
Besides, we aggregate both the parameters and the return value of a function into a cohesive \textit{function entity}. 
The omission of these objects allows our analysis to focus computational resources on the dependencies that matter most for repository-level reasoning.

\subsubsection{Inter-Entity Dependencies}
\label{sec:ied}
Based on the concept of entities, we lift type dependencies from the level of variable occurrences to the level of entities. 
Formally, an entity $E_1$ \textit{depends} on another entity $E_2$ if and only if there exists one variable occurrence $o_1$ within the syntactic scope of $E_1$ and another occurrence $o_2$ within $E_2$, such that $o_1$ is dependent on $o_2$. 

\begin{figure}[t!]
    \centering
    \includegraphics[width=\textwidth]{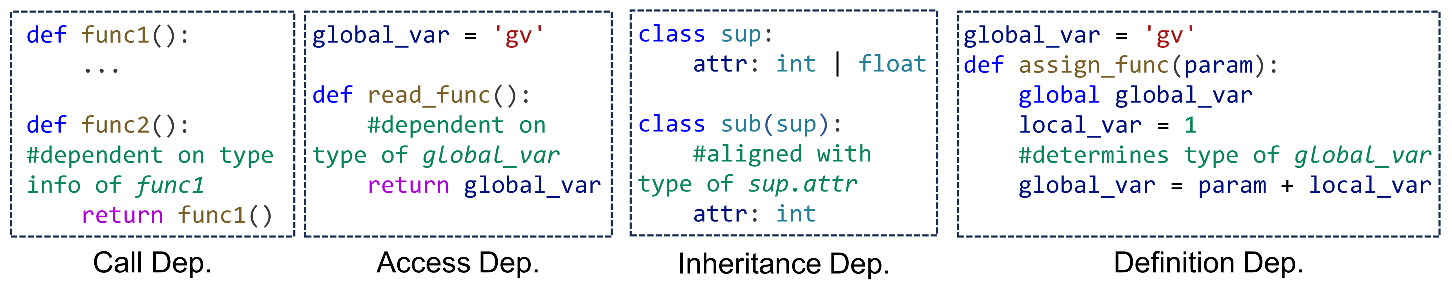}
    \vspace{-2em}
    \caption{Examples of Dependency Patterns in \methodName}
    \label{fig:dep_examples}
    \vspace{-1.5em}
\end{figure}

However, precisely tracking all such dependencies via deep inter-procedural static analysis is computationally expensive and often intractable, especially in a dynamic language like Python. 
To address this, \methodName~ employs a lightweight, pattern-based approach to efficiently identify critical inter-entity dependencies, as exemplified in Fig.~\ref{fig:dep_examples}. We define four primary dependency patterns:
\begin{itemize}
    \item \textbf{Call Dependency (Variable/Function $\rightarrow$ Function)}. If a function entity $f$ is called during the definition of another variable or function entity $e$, \methodName~establishes a dependency edge $e \rightarrow f$. The type of f's return value influences the type of the variable in $e$ that receives this value. Moreover, the types of arguments passed from $e$ to $f$ should be aligned with the types of $f$'s parameters.
    \item \textbf{Access Dependency (Variable/Function $\rightarrow$ Variable)}. An entity $e$ is dependent on a variable entity $v$ if $e$ accesses the value of $v$. The type of $v$ therefore constrains the type of the variable within $e$ that holds the accessed value.
    \item \textbf{Inheritance Dependency (Class $\rightarrow$ Class)}. This dependency arises from object-oriented inheritance. A subclass $C_{sub}$ is dependent on its parent class $C_{super}$. The type of a method or attribute in $C_{sub}$ must be compatible with its declaration in $C_{super}$.
    \item \textbf{Definition Dependency (Variable $\rightarrow$ Function)}.  A variable entity $v$ is dependent on a function entity $f$ if $f$ assigns a value to $v$. The type of the value being assigned within $f$, i.e., the right-hand side of the assignment, constrains the type of $v$.
\end{itemize}

The inaccuracies introduced by our pattern-based extraction are manageable.
First, superfluous edges do not significantly impede the inference capabilities of the LLM, as its attention mechanism allows it to selectively focus on the most relevant information during type resolution. Second, missing dependencies are designed to be progressively identified and incorporated during the subsequent EDG refinement phase.

\subsubsection{EDG Construction}

\methodName~employs static analysis on Abstract Syntax Trees (ASTs) to systematically extract entities and inter-entity dependencies from the input repository and construct the initial EDG. 
Specifically, \methodName~identifies entities based on the type of AST node.
Furthermore, \methodName~analyzes each statement in the definition code to build dependency edges in the EDG. 
\methodName~determines the category of dependency, according to the pattern defined in Sec.~\ref{sec:ied}, based on the AST node type and scope, and utilizes a static analysis tool PyAnalyzer~\cite{jin2024pyanalyzer} to locate objects accessed within statements.
  
To illustrate the construction process of EDG, we revisit the motivating example of a type inference task with complex constraints, as depicted in Fig.~\ref{fig:complex_dep}. 
This code snippet involves four entities: the function entity \texttt{after\_this\_request}, the variable entities \texttt{\_cv\_request} and \texttt{RequestContext.\_after\_request\_function}, and the class entity \texttt{RequestContext}. 
Among the two inter-procedural dependencies \textcircled{6} and \textcircled{7}, the usage-definition dependency, labeled as \textcircled{6}, is successfully identified by matching the \textit{Access Dependency} pattern. 
The complex dependency \textcircled{7} cannot be resolved at this stage and will be progressively incorporated in subsequent iterations.

\subsection{Phase I: Entity Reorganization}


To construct a compact and highly relevant context for type inference, \methodName~ leverages the type annotations of dependent entities, rather than their full definition code or deeper dependencies. Processing these dependencies naturally calls for a topological ordering of the EDG. 
However, mutual dependencies create cycles, making a standard topological sort impossible. 
\methodName~ overcomes this by systematically identifying and collapsing these cycles into entity clusters. 
This core transformation reduces the EDG to a Directed Acyclic Graph (DAG) of entity clusters, enabling \methodName~ to finally process them in a valid topological order for inference.

\subsubsection{Aggregating Entities}

\methodName~transforms the EDG into a DAG by aggregating mutually dependent entities into one entity cluster.
These mutually dependent entities manifest as directed cycles within the EDG. 
Crucially, since cycles may share overlapping vertices, each maximal strongly connected component (SCC) represents a set of mutually dependent entities where any pair exhibits cyclic dependency. 
Therefore, \methodName~ leverages the APIs provided by the python library  \textit{networkx}~\cite{hagberg2020networkx} to
collapses such SCCs into atomic Entity Clusters, transforming the EDG into a DAG of clusters that satisfy partial ordering constraints.

Moreover, constrained by LLM context window limitations, the size of an entity cluster must be limited.
\methodName~enforces an entity cluster cardinality bound of five entities by selectively removing dependency edges.
To mitigate the potential loss of inference accuracy caused by edge removal, \methodName~leverages a greedy algorithm to remove dependencies inside the cluster and decompose it into small ones.
Specifically, \methodName~iteratively removes dependency edges until the entity cluster is partitioned into multiple clusters that satisfy the cardinality constraint. 
In each iteration, \methodName~identifies an edge  it searches for an edge such that its removal decomposes this SCC into multiple strongly connected components and proceeds to remove that edge. 
This pragmatic trade-off marginally reduces inference accuracy while preserving critical dependency relationships, maintaining computational feasibility.


\subsubsection{Identifying Target Entity Clusters}
\label{sec:identify_target}
As described above, \methodName~leverages the type annotations of dependent entities to supply essential dependent type information.
Therefore, in each iteration, only entity clusters where all dependent entities have been annotated can serve as the target for type inference, allowing us to ignore other entity clusters to conserve computational resources.
Consequently, after partitioning the EDG into entity clusters, \methodName~ removes all entities that contain type annotations from the generated DAG and extracts the entity clusters with in-degree zero from the remaining graph as the objectives for type inference in the current iteration. 
Since the DAG is acyclic, entity clusters with in-degree zero are guaranteed to exist in the residual graph, thus ensuring that the target entity clusters can be reliably extracted in every iteration.

\subsection{Phase II: EDG Refinement}

In this phase, \methodName~leverages LLMs to refine both the dependency edges and type annotations of entities.
The underlying idea is that, since LLMs are capable of inferring correct types when provided with sufficient contextual information, they can also identify the informational gaps that hinder the completion of the inference process. 
Based on this idea, after constructing the type-inference context for each target entity cluster,  \methodName~sequentially performs dependency completion and type inference using this context, thereby achieving a co-evolution of type-related dependencies and type annotations.

\subsubsection{Identifying Missing Dependencies}

\methodName~leverages LLMs to identify the lacking edges in the EDG. 
Specifically, \methodName~constructs the prompt to identify the lacking dependencies for each entity cluster based on its type-inference context, including its definition code and dependencies.
The primary function of these prompts is to instruct the LLM to assess whether the current type-inference context is adequate. Furthermore, they guide the LLM in pinpointing any absent dependencies that may hinder the completion of the inference process. 

Consider the function entity \texttt{after\_this\_request} in Fig.~\ref{fig:complex_dep}. 
With the type-inference context containing its definition code and the type of dependent global variable \texttt{\_cv\_request},  the LLM can determine that the reference of variable \texttt{\_cv\_request}  is of type \texttt{ContextVar[RequestContext]}. 
Leveraging its knowledge of the Python standard library, particularly the \texttt{ContextVar} class, the LLM can further deduce that the local variable \texttt{ctx} is of type \texttt{RequestContext}.
Based on these insights, the LLM identifies a critical gap: the type of \texttt{RequestContext.\_after\_request\_functions} remains unspecified in the type-inference context. 
Consequently, \methodName~identifies the absence of the inter-procedural dependency \textcircled{7}, which is crucial for resolving the type-inference task.

\subsubsection{Generating Type Annotations}
Since the missing dependencies are identified based on the latent EDG, the entity clusters that are detected to have missing dependencies may contain additional missing dependencies, which can only be identified after the EDG is updated. 
Therefore, \methodName~performs type inference exclusively on the entity clusters without missing dependencies. 
For each entity cluster with complete dependency, \methodName~constructs the prompt by utilizing the type-inference context of the entity cluster and a predefined prompt template, and queries LLMs to generate the type annotations.

\subsection{Phase III: Type Validation}

The final phase of the iteration is dedicated to ensuring the correctness and consistency of the generated type annotations. 
It operates through a two-step process: first, detecting type conflicts using a static type checker, and second, employing a backtracing approach to correct the identified errors. 
The ultimate objective is to produce a set of \textit{conflict-free} type annotations that can be successfully validated by standard Python type checkers.

\subsubsection{Detecting Type Conflict}

Erroneous type annotations can propagate through dependency chains, leading to a cascade of incorrect inferences for downstream entities. 
To mitigate this risk, \methodName~employs the static type checker Mypy~\cite{lupuliak2025static} to identify the erroneous type annotations. 
The validation process operates on a working copy of the original unannotated source code to isolate the effects of new annotations. 
Specifically, whenever a type annotation is generated for an entity, \methodName~applies it to this copy. 
It then invokes Mypy to perform a static analysis on the modified codebase. If Mypy reports any type errors, \methodName~flags the newly added annotation as the source of a \textit{type conflict}, and the annotation is immediately reverted from the working copy.
This incremental and cautious approach prevents faulty annotations from corrupting the type environment, ensuring that each validation step occurs in a consistent context.

\subsubsection{Correcting Error Types}

Upon detecting a type conflict, \methodName~employs a backtracking strategy to restore type consistency.
The procedure retracts the type annotation that caused the conflict and triggers a re-inference process to find a compatible alternative.
For most conflicts, \methodName~ considers the most recently inferred type to be the source of the conflict. 
Consequently, this new type is discarded, allowing \methodName~ to derive a more suitable type.
However, in conflicts between the type of  and corresponding parameter in the function definition and between the type of an attribute in a superclass and subclass, the error can originate from either side of the conflict, compelling \methodName~to apply a more sophisticated resolution strategy.

Function parameters present a distinct challenge, as their correct type often depends on usage patterns across the entire codebase. 
\methodName~handles two primary scenarios. 
If an overly permissive parameter type leads to invalid operations within the function body (e.g., calling \texttt{append()} on a variable typed as \texttt{str}), \methodName~provides the specific Mypy error diagnostic as feedback to the LLM, prompting it to perform iterative type narrowing. Conversely, if an overly restrictive parameter type causes type errors at its call sites, the annotation is deemed fundamentally flawed. This triggers an invalidation of the function's annotation. The function is then re-queued for inference, now augmented with the specific conflict context to guide a more accurate, corrective regeneration.

Inheritance hierarchies introduce another layer of complexity, particularly concerning method signature mismatches that violate the Liskov Substitution Principle. 
For instance, a child method might have a parameter type that is not a valid contravariant of the parent's. 
To resolve such discrepancies, \methodName~invalidates the annotation of the parent method. The parent method is then submitted for re-inference, with the child's method signature provided as a crucial constraint. 
This ensures the newly generated parent annotation is compatible, allowing corrections to propagate up the inheritance chain and restore subtyping consistency.

For all other categories of type errors, \methodName~employs a direct feedback loop. The raw error messages from Mypy are fed back to the LLM, which is prompted to generate a refined annotation. This iterative cycle of verification and repair continues until the annotation becomes conflict-free or a predefined iteration threshold is exceeded.

If \methodName~fails to find the suitable type, as a final fallback for persistent conflicts, \methodName~adopts a conservative strategy: it replaces the problematic annotation with the \texttt{Any} type. 
While this sacrifices some type precision, it guarantees the elimination of the type conflict, ensuring that the final output is a fully valid and conflict-free set of annotations.

\section{Evaluation}

To evaluate the effectiveness of \methodName, we investigate the following research questions (RQs):

\begin{itemize}
    \item   \textbf{RQ1 (Type Accuracy)}: How accurately do the inferred type annotations align with ground truth types?
    \item  \textbf{RQ2 (Global Consistency)}: To what extent do generated annotations maintain repository-level consistency across dynamic call resolutions and inheritance hierarchies?
     \item  \textbf{RQ3 (Ablation Study)}: What is the contribution of each core component to overall system performance?
     \item  \textbf{RQ4 (Efficiency)}: What is the time cost for type inference?
\end{itemize}

The machine for running experiments is equipped with an Intel Xeon Gold 6330 CPU (28 cores, 2.00GHz), Ubuntu 24.04, and Python 3.10.

\subsection{Evaluation Setup}
Before the evaluation, we introduce the implementation settings of our approach  and the construction of our benchmark dataset, as well as the baselines and metrics used in the evaluation.
 
\subsubsection{Implementation}
\methodName~is implemented in 5,000 lines of Python code. 
During EDG initialization, we leverage PyAnalyzer~\cite{jin2024pyanalyzer}, a SOTA static analysis tool, to precisely identify objects used in each statement, according to which we develop the dependency between entities.
To address the context window length limitations of LLMs, we set the maximum entity cluster size to 5 based on the entity lengths observed in our experiments. We allow the LLM to attempt inference on each entity up to three times, stopping when it produces a valid type annotation or when the attempt limit is reached.
To comprehensively evaluate our approach, in our ablation study, we employ both a generative model (\textit{DeepSeek-V3}) and a reasoning model (\textit{DeepSeek-R1}), which represent distinct architectural paradigms for type inference benchmarking. 
For other RQs, we utilize the results derived from \textit{DeepSeek-V3} to represent the performance of \methodName.

\subsubsection{Benchmark Construction}

\begin{table}[htbp]
  \vspace{-12pt}
  \centering
  \caption{The statistics of all repositories in benchmarks}
  \vspace{-10pt}
  \resizebox{0.95\columnwidth}{!}{
    \begin{tabular}{
      l
      S[table-format=4]
      S[table-format=4]
      S[table-format=4]
      S[table-format=4]
      S[table-format=4]
      S[table-format=4]
      r
    }
    \toprule
    \makecell{Repository} & {\#Files} & {\#Lines} & {\#Tokens} & {\#Vars} & {\#Funcs} & {\#Entities} & {Domain} \\
    \midrule
        pre\_commit\_hooks 5.0.0   & 35  & 1501  & 11672  & 62    & 105   & 163   & Code Quality  \\
        flake8         7.3.0   & 33  & 2848  & 18452  & 96    & 210   & 498   & Code Quality \\
        typer          0.16.0  & 16  & 2156  & 21653  & 133   & 164   & 441   & CLI\textsuperscript{*} \\
        flask          3.1.1   & 24  & 5645  & 27134  & 110   & 369   & 654   & Web Development \\
        pre\_commit    4.2.0   & 66  & 3816  & 34709  & 236   & 352   & 580   & Code Quality  \\
        fastapi        0.116.1 & 44  & 3742  & 35485  & 137   & 229   & 767   & Web Development  \\
        click          8.2.1   & 16  & 6628  & 39319  & 90    & 526   & 879   & CLI\textsuperscript{*} \\
        urllib3        2.5.0   & 36  & 6750  & 41774  & 199   & 492   & 1118  & Network Request \\
        black          25.1.0  & 25  & 6875  & 54479  & 124   & 393   & 681   & Code Formatting  \\
        jinja2         3.1.6   & 25  & 9074  & 57431  & 166   & 743   & 1495  & Web Development  \\
        werkzeug       3.1.3   & 52  & 13622 & 77033  & 161   & 1113  & 1964  & Web Development  \\
        rich           14.0.0  & 78  & 12819 & 115536 & 372   & 901   & 1980  & CLI\textsuperscript{*} \\
    \bottomrule
    \multicolumn{4}{l}{\small * CLI: Command-Line Interface}
    \end{tabular}%
  }
  \label{tab:repo_stats}%
  \vspace{-1.5em}
\end{table}

\newcommand{\RepoGroundTruth}{$R_{truth}$}
\newcommand{\RepoNoAnno}{$R_{\overline{anno}}$}
\newcommand{\RepoNoAnnoErrFree}{$R_{raw}$}
\newcommand{\RepoInput}{\RepoNoAnnoErrFree}
\newcommand{\RepoOutput}{$R_{out}$}

To evaluate \methodName's effectiveness, we construct a high-quality benchmark by selecting 12 projects from TypyBench~\cite{jiang2025typybench} that meet stringent quality criteria: GitHub stars exceeding 10,000 and type annotation coverage surpassing 85\%. 
To maintain the temporal validity of evaluation, we clone the repositories of these projects, and checkout to their latest release version on PyPI~\cite{index2021pypi}.
These repositories preserve developer-authored type annotations, serving as ground truth (\RepoGroundTruth) to assess type accuracy.
Table~\ref{tab:repo_stats} summarizes the characteristics of repositories, including size, entities that require type annotation, and domain.

Following TypyBench, we systematically removed all type annotations and type-related docstrings from \RepoGroundTruth~ to produce the unannotated repository \RepoNoAnno. To accurately evaluate the global type consistency in RQ2, we then performed prerequisite error suppression on \RepoNoAnno: we ran the static type checker Mypy to detect inherent type errors and, following PEP 484, added `\texttt{\#type: ignore}' comments to the erroneous statements. This yields an unannotated and error-free (i.e., globally consistent) initial repository \RepoNoAnnoErrFree, which we use to compare the number of type errors introduced in \RepoOutput~ after inferring types with \methodName. Notably, we configure Mypy to exclude \texttt{var-annotated}, \texttt{assignment}, and \texttt{has-type} error categories, as these stem from incomplete annotations rather than semantic defects, and suppress \texttt{missing-imports} errors since they are attributable to third-party dependencies.

\subsubsection{Evaluation Metrics}
Following TypyBench's methodology, we evaluate \methodName's performance through two complementary dimensions: type accuracy and global consistency.

\textbf{Type Accuracy Assessment}. 
We evaluate the accuracy of the type annotations produced by \methodName~ by computing two metrics, \textit{TypeSim} and \textit{TypeExact}, between the type-annotated repository \RepoOutput~ generated by \methodName~ and the ground-truth repository \RepoGroundTruth~.

\begin{itemize}
    \item \textit{TypeSim} measures functional similarity between inferred and ground truth types (i.e., types in \RepoOutput~ and \RepoGroundTruth) based on their shared methods and operations. 
    The similarity score of predicted type $T_p$ and ground truth type $T_g$ is computed as:
  \[
  \text{sim}(T_p, T_g) = \frac{| attrs(T_p) \cap attrs(T_g) |}{| attrs(T_p) \cup attrs(T_g) |}
  \]
  where $attrs(T)$ denotes the set of methods and operations supported by type $T$, excluding those common to all types that inherit from \texttt{object}. 
  This structural approach prioritizes behavioral equivalence over nominal matching, accommodating Python's duck typing paradigm.
  \item \textit{TypeExact} provides strict nominal accuracy, requiring exact type identity matches. This serves as a conservative baseline for traditional typing system.
\end{itemize}

\textbf{Global Consistency Measurement}.
\methodName~ processes \RepoInput~ and yields an annotated repository \RepoOutput. We quantify the degradation in global consistency by executing Mypy on \RepoOutput~ and recording the number of reported errors.
This metric directly reflects practical barriers developers would encounter when integrating inferred types.
For example, a function returning \texttt{List[int]} but consumed as \texttt{List[str]} would disrupt static type checking, undermining core benefits of type annotations. 
Such inconsistencies manifest tangible maintenance and refactoring obstacles.



\subsubsection{Baselines and Ablation Methods}

To the best of our knowledge, \methodName~is the first approach targeting \textit{repository-level} type inference. 
Consequently, we selected four SOTA \textit{function-level} type inference methods for comparative analysis. 
Specifically, for RQ1 and RQ2, we employed two \textit{SA-based} methods and two \textit{learning-based} methods as baselines.

\textbf{HiTyper}\cite{peng_static_2022} is a hybrid type inference approach for Python that integrates static analysis with deep learning. 
It constructs Type Dependency Graphs to model interdependencies among variables within a function body, enabling iterative type propagation. 

\textbf{QuAC}\cite{wu_quac_2024} employs an attribute-centric methodology.
It extracts attribute sets (e.g., accessed methods/properties) from expressions and infers types via information retrieval techniques. 
This is achieved by matching observed attributes against type-specific attribute sets using BM25 ranking.

\textbf{TypeGen}\cite{peng_generative_2023} is a generative type inference framework that combines static analysis with few-shot learning. 
It generates Chain-of-Thought prompts from code slices and then employs in-context learning with LLMs to predict types.

\textbf{TIGER}\cite{wang_tiger_2024} introduces a generating-then-ranking framework. 
It fine-tunes encoder-decoder models with contrastive learning, a generative model produces candidates, and a similarity model ranks them against user-defined types.

For RQ3, we conduct a comprehensive evaluation to assess the impact of \methodName's two core strategies: EDG-based dependency extraction and backtracking-based type conflict resolution, as well as the integration of LLM.
To quantify their contributions, we designed four ablation variants.

\textbf{$\methodName^{Base}$} where all strategies are removed. Entity definitions are directly fed to the LLM without dependency information or conflict resolution.

\textbf{$\methodName^{W/oDep}$} removes EDG-based dependency extraction. Provides only entity definitions to the LLM while retaining conflict resolution.

\textbf{$\methodName^{W/oFix}$} removes backtracking-based conflict resolution. Provides full dependency but skips type conflict correction.

\textbf{$\methodName^{R1}$} replaces the LLM used in \methodName, DeepSeek-V3, with a reasoning model DeepSeek-R1.

\begin{figure}[b!]
    \vspace{-1em}
    \centering
    \includegraphics[width=\textwidth]{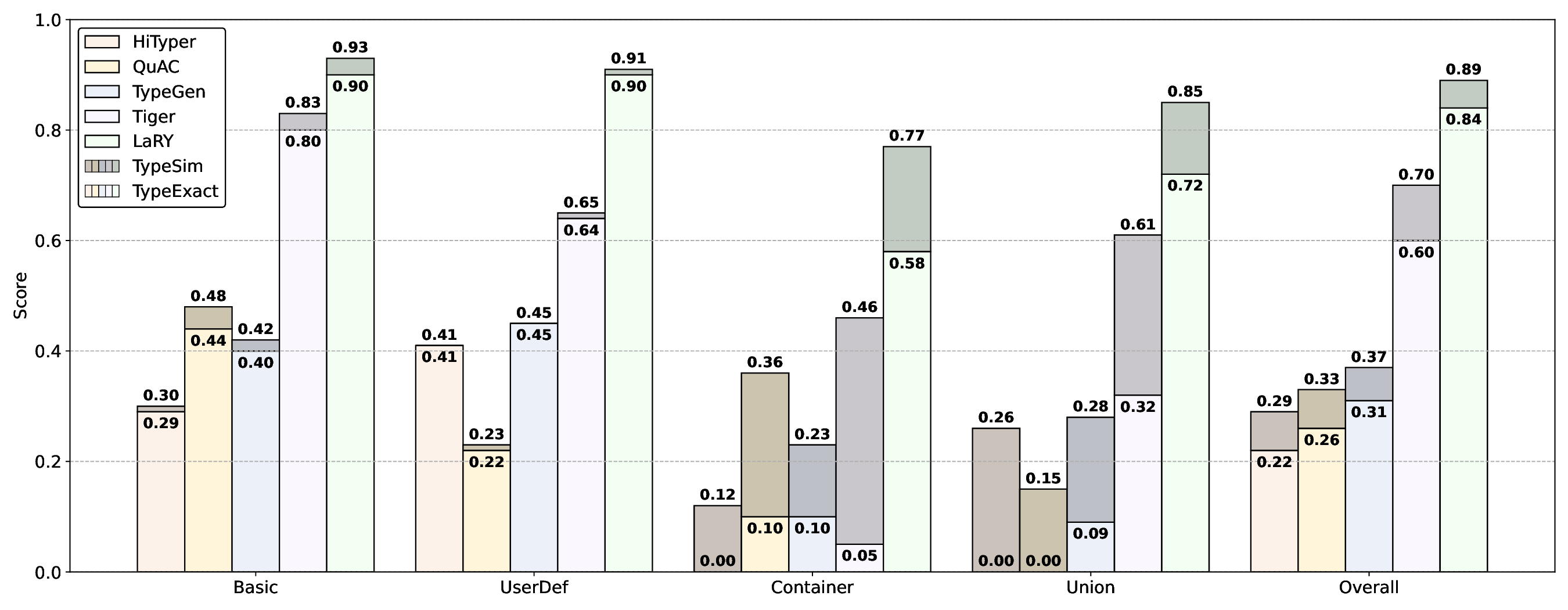}
    \vspace{-2em}
    \caption{Accuracy of \methodName~and baseline methods on inference of different categories of types}
    \label{fig:RQ1}
    \vspace{-1em}
\end{figure}

\subsection{RQ1. Type Accuracy}
To conduct a granular analysis of \methodName's accuracy across different types, we categorize the types used in benchmarks into five distinct groups: basic types (\texttt{int}, \texttt{float}, \texttt{str}, \texttt{bytes}, \texttt{None}), container types, Union types, user-defined types, and other complex types. 
The type prediction accuracy of \methodName~and baseline methods across these categories is shown in Fig.~\ref{fig:RQ1}.
The results  demonstrate the superior performance of \methodName. 
In the overall evaluation, \methodName~achieves a \textit{TypeSim} score of 0.89, representing a significant improvement of 27.1\% over the best-performing baseline, TIGER (0.70). Moreover, \methodName's exact match accuracy alone is 0.84, a score that surpasses all other methods. 
To better isolate the contribution of our proposed components, we consider the native LLM as a degenerative variant of our approach, $\methodName^{Base}$, and thus present this comparison between \methodName~and the native LLM in the ablation study in Sec. \ref{sec:ablation}.

The \textit{SA-based} methods, HiTyper and QuAC, perform reasonably well on Basic types (0.48 and 0.42, respectively), where direct data-flow analysis is effective. However, they struggle significantly with more ambiguous or complex scenarios like Union types and nested Container types, which often require a deeper semantic understanding of developers' intent.
In contrast, the \textit{learning-based} methods, TIGER and TypeGen, generally outperform the static analysis tools, indicating that their models have learned to recognize semantic patterns beyond what rule-based analysis can capture. 
TIGER, as the strongest baseline, shows respectable performance on UserDef (0.65) and Union (0.61) types. However, their one-shot prediction process, lacking a verification loop, still falls significantly short of \mbox{\methodName}'s accuracy. 

\begin{tcolorbox}[boxsep=4pt,left=0pt,right=0pt,top=1pt,bottom=0pt]
\noindent\textbf{Answer to RQ1:} \methodName~significantly outperforms the SOTA type inference methods in terms of accuracy. It achieves an overall accuracy of 0.89, with an exact match rate of 0.84. 
Its strength is especially evident in handling complex, user-defined, and composite types, demonstrating the effectiveness of its LLM-based generation coupled with a rigorous validation mechanism.
\end{tcolorbox}

\subsection{RQ2. Global Consistency}

The errors introduced by \methodName~and baseline methods are presented in Table~\ref{tab:type_errors_introduced}.
The results clearly demonstrate the  global consistency of the type annotations produced by \methodName. 
In total, \methodName~introduces only 554 type errors across all repositories, a number substantially lower than all baseline methods. 
Specifically, \methodName~reduces the number of introduced type errors by 79.1\% compared to the best-performing baseline, HiTyper (2,654 errors), and by up to 93.5\% compared to TIGER (8,577 errors). 
This significant reduction underscores \methodName's effectiveness in generating annotations that are not just locally plausible but also globally coherent.
Similar to RQ1, the comparison with the native LLM is presented in our ablation study in Sec. \ref{sec:ablation}.

\begin{table}[h!]
  \vspace{-1em}
  \centering
  \caption{Type errors introduced by adding inferred type annotations (Aggregated over 12 repositories). Percentages represent the proportion of each error type relative to the total errors introduced by each method.}
  \label{tab:type_errors_introduced} 
  \resizebox{0.92\columnwidth}{!}{
  \begin{tabular}{l r@{ }l r@{ }l r@{ }l r@{ }l r@{ }l}
    \toprule
    
    \makecell{Error Type} & \multicolumn{2}{c}{HiTyper} & \multicolumn{2}{c}{QuAC} & \multicolumn{2}{c}{TypeGen} & \multicolumn{2}{c}{Tiger} & \multicolumn{2}{c}{\methodName} \\
    
    \midrule
    arg-type                   & 671 & (25.3\%) & 1,047 & (28.1\%) & 2,923 & (34.1\%) & 1,015 & (19.5\%) &   0 & (0.0\%) \\
    assignment                 & 403 & (15.2\%) &   849 & (22.8\%) & 1,126 & (13.1\%) &   778 & (15.0\%) & 543 & (98.0\%) \\
    attr-defined               & 233 & (8.8\%)  &   547 & (14.7\%) &   506 & (5.9\%)  &   366 & (7.0\%)  &   0 & (0.0\%) \\
    return-value               & 325 & (12.2\%) &   385 & (10.3\%) &   616 & (7.2\%)  &   190 & (3.7\%)  &   0 & (0.0\%) \\
    call-arg                   & 215 & (8.1\%)  &   109 & (2.9\%)  &   289 & (3.4\%)  &   283 & (5.4\%)  &   0 & (0.0\%) \\
    override                   & 104 & (3.9\%)  &   118 & (3.2\%)  &   348 & (4.1\%)  &   142 & (2.7\%)  &   0 & (0.0\%) \\
    var-annotated              &  90 & (3.4\%)  &    73 & (2.0\%)  &    78 & (0.9\%)  &    77 & (1.5\%)  &   6 & (1.1\%) \\
    name-defined               &   7 & (0.3\%)  &    74 & (2.0\%)  & 1,217 & (14.2\%) & 1,496 & (28.8\%) &   0 & (0.0\%) \\
    \textit{Other Errors}\textsuperscript{*} & 606 & (22.8\%) & 520 & (14.0\%) & 1,474 & (17.2\%) & 855 & (16.4\%) & 5 & (0.9\%) \\
    \midrule
    \textbf{Total Errors}      & \multicolumn{2}{c}{\textbf{2,654}} & \multicolumn{2}{c}{\textbf{3,722}} & \multicolumn{2}{c}{\textbf{5,202}} & \multicolumn{2}{c}{\textbf{8,577}} & \multicolumn{2}{c}{\textbf{554}} \\
    \bottomrule
    \\[-1em]
    \multicolumn{11}{l}{\footnotesize \textsuperscript{*} Includes less frequent error types such as `\texttt{operator}', `\texttt{index}', `\texttt{list-item}', `\texttt{misc}', etc., detected using Mypy.}\\
  \end{tabular}
  }
  \vspace{-1em}
\end{table}

A detailed analysis of the error types reveals the inherent challenges of achieving global consistency in automated type inference. 
A primary challenge is maintaining \textit{inter-procedural consistency}, as a type annotation for a single function must be compatible with every call site across the entire repository. 
This difficulty is reflected in the high number of arg-type errors observed in baseline methods (ranging from 19.5\% to 34.1\% of their totals), which often arise when a locally-inferred parameter type conflicts with the type of an argument passed from a different module. 
In contrast, \methodName~models inter-procedural dependencies using the EDG and employs a type-dependency co-evolution iteration to extract these dependencies.
Moreover, for learning-based methods like TypeGen and Tiger, without a mechanism to ensure that a generated type name is valid and accessible in its scope, LLMs may produce plausible types, leading to a large volume of name-defined errors. 
While \methodName~also leverages an LLM, its validation phase immediately detects and corrects such invalid types.
This proactive analysis and reactive validation work in concert, enabling \methodName~to eliminate arg-type, attribute-defined, return-value, and name-defined errors by design.

The errors introduced by \methodName~consist of assignment and var-annotated errors.
These two categories of errors are primarily triggered by the absence of type annotations. 
Consequently, \methodName~ omits these errors during iterative refinement to prevent excessive noise in the validation phase.
These errors reveal the current limitations of \methodName's static analysis. 
However, the introduced errors are few and represent missing or incomplete information rather than the introduction of actively incorrect types. 
Therefore, the impact of these errors is contained. 

\begin{tcolorbox}[boxsep=4pt,left=0pt,right=0pt,top=1pt,bottom=0pt]
\noindent\textbf{Answer to RQ2:} \methodName~generates significantly more globally consistent type annotations than SOTA baselines. Its novel validation phase, which iteratively checks and corrects annotations using a live type checker, allows it to effectively eliminate common cross-file and inter-procedural type conflicts that plague other methods. This proactive approach reduces the number of \texttt{Mypy} errors introduced by 79.1\% to 93.5\%, thereby minimizing the subsequent manual effort required by developers to fix inconsistencies after automatic type annotation.
\end{tcolorbox}

\subsection{RQ3. Ablation Study}
\label{sec:ablation}

As shown in Table~\ref{tab:RQ3}, the baseline variant, $\methodName^{Base}$, representing the native LLM, performs the poorest across all metrics. 
It achieves a TypeExact score of only 0.72 and introduces a substantial 4,186 static errors. 
This result highlights the inherent limitation of prompting an LLM with isolated code snippets, as it fails to comprehend the intricate web of inter-procedural dependencies, leading to lower accuracy and widespread type inconsistencies.

\begin{table}[h!]
  \centering
  \caption{Performance Comparison of \methodName~Variants}
  \vspace{-0.5em}
  \scalebox{0.8}{
    \begin{tabular}{lrrrrr}
    
    \toprule
          & $\methodName^{Base}$ & $\methodName^{W/oFix}$ & $\methodName^{W/oDep}$ & $\methodName^{R1}$ & \methodName \\
    \midrule
    TypeSim & \cellcolor[rgb]{ 1,  .78,  .808} 0.81  & 0.88  & 0.87  & \cellcolor[rgb]{ .776,  .937,  .808} 0.89  & \cellcolor[rgb]{ .776,  .937,  .808} 0.89  \\
    TypeExact & \cellcolor[rgb]{ 1,  .78,  .808} 0.72  & 0.81  & 0.81  & \cellcolor[rgb]{ .776,  .937,  .808} 0.83  & \cellcolor[rgb]{ .776,  .937,  .808} 0.84  \\
    Introduced Errors & \cellcolor[rgb]{ 1,  .78,  .808} 4,186  & 3,478  & 659   & \cellcolor[rgb]{ .776,  .937,  .808} 335  & \cellcolor[rgb]{ .776,  .937,  .808} 554  \\
    \bottomrule
    \end{tabular}%
    }
  \label{tab:RQ3}%
  \vspace{-0.5em}
\end{table}

When we introduce only the backtracking-based conflict resolution mechanism ($\methodName^{W/oDep}$), performance improves significantly. 
TypeExact increases by 9 percentage points to 0.81, and the number of introduced errors plummets by 84.2\% from 4,186 to 659. 
This demonstrates that even without explicit dependency information, integrating a static type checker into the inference loop allows \methodName~to iteratively correct localized errors, thereby enhancing both accuracy and type safety. 
However, the remaining 659 errors indicate that without a global understanding of dependencies, the model's ability to resolve complex, non-local conflicts is limited.
Conversely, the $\methodName^{W/oFix}$ variant, which incorporates dependency acquisition but omits the conflict resolution mechanism, reveals a critical insight. 
While providing dependency context boosts TypeExact accuracy to 0.81 and TypeSim to 0.88, it results in 3,478 introduced errors. 

The full \methodName, which combines both EDG-based dependency extraction and backtracking-based conflict resolution, achieves the highest performance. It outperforms all variants, reaching a TypeExact score of 0.84 and a TypeSim score of 0.89, and reducing introduced errors to 554. 

Finally, we examine the impact of the LLM choice by comparing the full \methodName~(using DeepSeek-V3) with the $\methodName^{R1}$ variant (using DeepSeek-R1). 
$\methodName^{R1}$ achieves a TypeExact score of 0.83, nearly identical to the 0.84 of the standard \methodName.
Moreover, it introduces significantly fewer static errors, reducing 40\% from 554 to 335.
This substantial improvement in type safety can be attributed to DeepSeek-R1's superior logical reasoning abilities.
A more powerful reasoning engine is better at understanding complex semantic relationships and maintaining global consistency, resulting in more reliable and correct type annotations. 
However, this enhanced reasoning capability comes at a higher computational cost. Therefore, the choice of LLM presents a practical trade-off between inference quality and efficiency.
Our default configuration of \methodName~with DeepSeek-V3 represents a balanced choice for general use, while a more powerful reasoning model like DeepSeek-R1 is a viable option for applications.

\begin{tcolorbox}[boxsep=4pt,left=0pt,right=0pt,top=1pt,bottom=0pt]
\noindent\textbf{Answer for RQ3:} Both EDG-based dependency extraction and backtracking-based conflict resolution are critical to \methodName's performance. They have a strong synergistic effect: dependency extraction provides essential context for accurate prediction, while conflict resolution is indispensable for maintaining global type consistency and preventing error propagation. The choice of LLM presents a practical trade-off between inference quality (fewer errors) and computational cost.
\end{tcolorbox}

\subsection{RQ4. Efficiency}

\begin{figure}[tbp]
    \centering
    \includegraphics[width=0.75\textwidth]{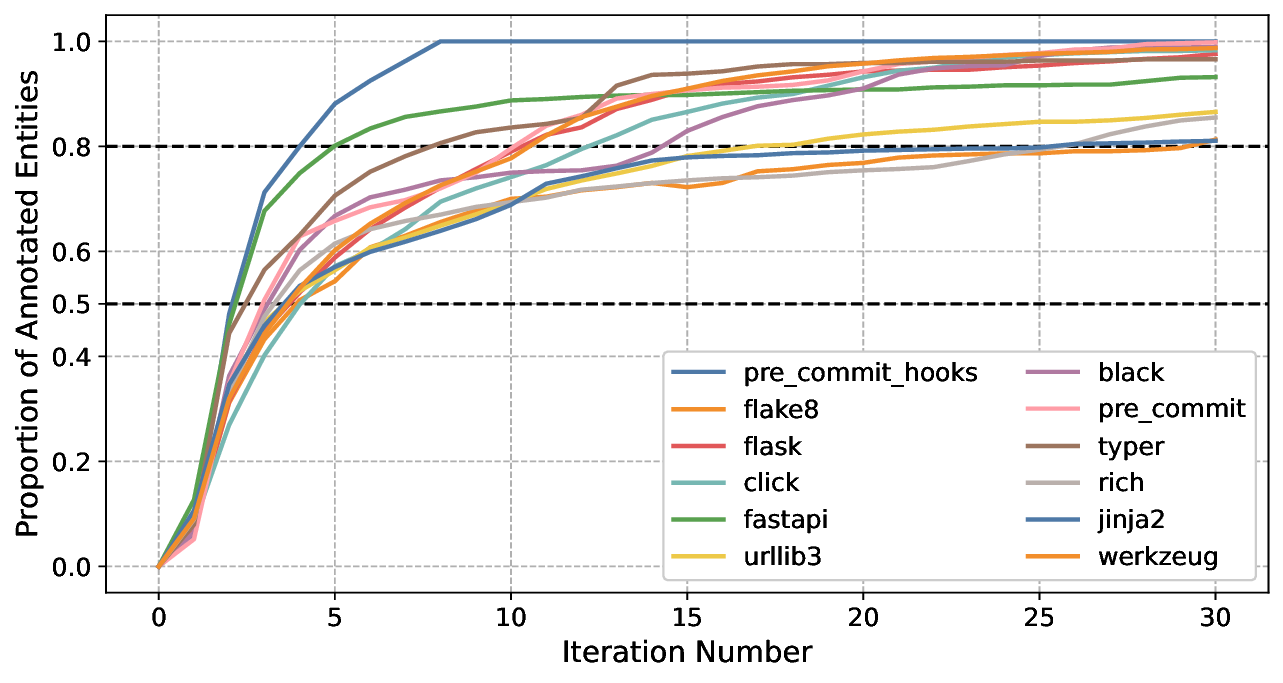}
    \vspace{-1em}
    \caption{Annotation rate over iterations across repositories }
    \label{fig:RQ4}
    \vspace{-1em}
\end{figure}

To evaluate the execution efficiency of \methodName, we analyze both the number of iterations and the time cost required for type inference on the benchmark.

As illustrated in Fig.~\ref{fig:RQ4}, \methodName~demonstrates remarkable efficiency across all 12 repositories. 
The type annotation process for each repository follows a characteristic pattern: an initial phase of rapid progress followed by a phase of gradual refinement. 
In the early stages, \methodName~swiftly completes type inference for entities with simple dependencies. 
The curves in the figure confirm this trend, showing that  \methodName~annotates over 50\% of entities within the first five iterations for all repositories. Subsequently, the rate of annotation decelerates as the system begins to resolve more complex, long-range dependencies between entities, requiring more feedback cycles.

The total number of iterations required for complete type inference varies with the scale and complexity of each repository.
For instance, \textit{pre\_commit\_hooks}, a smaller repository with only 163 entities, achieves complete annotation in under 10 iterations. 
In contrast, for larger and more complex repositories such as \textit{urllib3}, \textit{jinja2}, \textit{werkzeug}, and \textit{rich}, which each contain over 1,000 entities, \methodName~requires approximately 30 iterations to surpass 80\% annotation coverage. 
Despite these variations, the process demonstrates robust performance, consistently achieving over 80\% annotation coverage for all 12 repositories within 30 iterations.

Moreover, we measured the total execution time for each repository, with detailed results provided in the supplementary material. 
Across the benchmark, the average time to complete the type annotation process for a single repository is 3.56 hours. Similar to the iteration count, the execution time is correlated with the repository's characteristics. 
Empirically, we found that the annotation time is positively correlated with both the total number of tokens in the repository, which serves as a proxy for its scale, and the average number of parameters per function, which can indicate its code complexity.

\begin{tcolorbox}[boxsep=4pt,left=0pt,right=0pt,top=1pt,bottom=0pt]
\noindent\textbf{Answer to RQ4:} \mbox{\methodName}'s iterative type inference process is highly efficient. It rapidly annotates over 50\% of entities within the first five iterations and consistently achieves over 80\% coverage within 30 iterations across diverse repositories. The average execution time is 3.56 hours per repository, with the required time and number of iterations being positively correlated with the repository's scale and code complexity. This demonstrates that our approach is practical for real-world applications.
\end{tcolorbox}

\section{Threats to Validaty}

A primary threat concerns the scalability of \methodName. 
The computational cost of \methodName~is correlated to the size and complexity of the target repository. 
For large and complex repositories, the processing time might become unaffordable. 
To mitigate this threat, we have designed \methodName~to support an incremental, iterative workflow. 
Besides processing an entire codebase at once, \methodName~can also be applied on a per-module or per-commit basis during the software development lifecycle.

Another threat lies in the potential presence of pre-existing type errors within real-world repositories. 
Since \methodName~infers types based on the observed behavior and interactions of entities, type errors in the source code can mislead the inference process, leading to the generation of incorrect type annotations. 
To address this, during the construction of our evaluation dataset, we meticulously identified and masked potential static type errors.
This ensures that our experimental evaluation is conducted on a sanitized baseline, allowing for a fair assessment of \methodName's inference capabilities without confounding factors from legacy code issues. 
Moreover, within the \methodName~framework, our backtracking-based conflict resolution mechanism is designed to handle such situations. When an unresolvable type conflict is detected, \methodName~assigns the `\texttt{Any}' type, preventing the propagation of an error from one incorrect entity to its dependents. Furthermore, the explicit insertion of `\texttt{Any}' serves as a diagnostic signal for developers, flagging regions of the code where underlying type inconsistencies may exist and warrant manual inspection.

\section{Related Work}
\label{sec:related_work}

In this section, we provide a comprehensive review of the related work on Python type inference, including \textit{static-analysis-based} and \textit{learning-based} approaches. 

\subsection{Static-Analysis-based Type Inference}

A line of research in type inference leverages static analysis. These methods operate on a set of pre-defined heuristics and formal rules to deduce variable types~\cite{anderson2005towards,jensen2009type, hassan2018maxsmt,hassan2017smt}. By traversing a program's AST, they propagate type constraints throughout the codebase. This paradigm has given rise to several widely-adopted industrial tools, including Pyright/Pylance~\cite{pyright}, Pyre~\cite{Pyre}, pytype~\cite{pytype}, and the official Python type checker, Mypy~\cite{lupuliak2025static}. Concurrently, numerous academic proposals have explored static type inference not only for Python~\cite{xu2016python,pavlinovic2021data} but also for other dynamically-typed languages like JavaScript~\cite{chen2016principal,emmi2016symbolic}.

Despite their high precision in predictable code contexts, the efficacy of static analysis-based approaches is often constrained by the inherent dynamism of languages like Python. Features such as dynamic attribute access, metaprogramming, and calls to external, unannotated libraries present significant challenges~\cite{peng_static_2022}. Consequently, these methods frequently suffer from low coverage, failing to infer types in a substantial number of real-world scenarios.

\subsection{Learning-based Type Inference}

To address the challenges posed by Python's dynamic nature, a growing body of research has shifted towards learning-based methods, which learn to infer types from vast code corpora.

Early efforts in this domain framed type inference as a classification task, training supervised models to predict a type for a given variable from a predefined set of candidates. For instance, TypeWriter~\cite{pradel2020typewriter} combines probabilistic prediction with a search-based refinement process, validating its inferences with a gradual type checker. Lambdanet~\cite{weilambdanet} pioneered the use of graph neural networks (GNNs) on a program's type dependency graph to predict both standard and user-defined types in TypeScript. DLInfer~\cite{yan2023dlinfer} first extracts program slices for variables and then employs a Bi-directional GRU to model type propagation from these contexts. More recently, TypeBert~\cite{jesse2021learning} demonstrated that transformer-based models, with their strong inductive bias for token sequences, can achieve state-of-the-art performance when trained on sufficient data. A principal drawback of these classification-based models is their struggle with out-of-vocabulary (OOV) types. This severely limits their practical applicability in real-world projects where new user-defined and library types are prevalent.

The recent success of large pre-trained models for code~\cite{feng2020codebert,wang2021codet5,guo2022unixcoder,fried2022incoder} has inspired a new wave of research. This line of work reformulates type inference as a conditional text generation problem, often as a fill-in-the-blank task. A special placeholder is inserted at the annotation site, and the model is prompted to generate the most probable type based on the surrounding code context. To enhance performance, some studies fine-tune these models on type-annotated codebases~\cite{weitypet5}. 

More recently, the advent of LLMs has opened new avenues for type inference. TypeGen~\cite{peng_generative_2023} exemplifies this trend by utilizing powerful, parameter-frozen LLMs such as GPT-3.5 and GPT-4. It synergizes lightweight static analysis with in-context learning~\cite{dong2024survey} and Chain-of-Thought (CoT)~\cite{wei2022chain} prompting to generate not only type predictions but also their rationales. While demonstrating impressive performance, the reliance on proprietary, resource-intensive LLMs introduces significant practical hurdles, primarily concerning scalability and computational cost, which motivates our work.

\section{Conclusion}

In this paper, we propose \methodName, an LLM-powered approach to automatically generate type annotations for Python repositories. 
\methodName~employs a novel Entity Dependency Graph to model inter-procedural type dependencies in the repository. 
During each iteration, dependencies and type information co-evolve through mutual refinement. 
Experimental results show that \methodName's effectiveness outperforms baseline approaches across all the metrics, achieving higher type inference accuracy with fewer introduced type errors.

\section{Data Availability}

Our replication package is publicly available at \url{https://anonymous.4open.science/r/PyTIR-E304/}.

\bibliographystyle{ACM-Reference-Format}
\bibliography{acmart}

\end{document}